\documentstyle[11pt,aaspp4]{article}

\slugcomment{Accepted for publication in ApJ}

\lefthead{Sparks et al.}
\righthead{Disks and Jets}

\begin{document}

\title{Face-on dust disks in galaxies with optical jets\footnote{Based on observations with the NASA/ESA Hubble Space Telescope obtained at the Space Telescope Science Institute, which is operated by the Association of Universities for Research in Astronomy, Incorporated, under NASA contract NAS5-26555.}}

\author{William B. Sparks,
Stefi A. Baum,
John Biretta,
F. Duccio Macchetto\footnote{On assignment from the Space Science Department
of the European Space Agency.}}
\affil{Space Telescope Science Institute, 3700 San Martin Drive, Baltimore, MD 21218}

\author{Andr\'e R. Martel}
\affil{Dept. of Physics and Astronomy, The Johns Hopkins University, Baltimore, MD 21218}



\begin{abstract}
The presence of optical synchrotron jets in radio galaxies is relatively
rare.
Here, we show that of the nearest five FR-I 3CR radio galaxies showing optical jets,
{\it four} show evidence for almost circular, presumably face-on, dust disks.
This is strong support for the two-fold idea that (\romannumeral1)~jets emerge
close to perpendicular to inner gas disks and (\romannumeral2)~optical
non-thermal synchrotron emission is seen only when the jet points
towards the observer.
The implied critical angle to the line-of-sight
is approximately $30$ --- $40^{\circ}$,
i.e. if the angle of the jet to the line-of-sight
is less than about $40^{\circ}$\ we see an optical jet.
The corresponding relativisitic $\gamma$ factor is $\approx 1.5$
which is consistent with current observations of jet proper motion
that show a range up to $\gamma \sim 6$ for M87.
The relatively low speeds implied by $\gamma \approx 1.5$ may
be due to a global deceleration of the jet as in unified
theories, or else to stratification within the jet.
Unresolved nuclei are common in the optical. Their luminosities
are also consistent with the beaming concept when compared to inclination
inferred from the dust lanes.
The disk sizes are typically several hundred parsecs, to kiloparsec size.
The galaxy with an optical jet that does not show a face-on disk, M87,
instead has more complex radial dust and ionized gas filaments.

\end{abstract}




%

\section{Introduction}

The reasons why radio galaxies project high energy jets of
relativistic plasma over vast, galactic and even extragalactic,
distances from a tiny volume at the nucleus of the host
galaxy remain unknown. Essentially {\it all} radio galaxies seem to
display jets when observed with sufficient sensitivity and resolution,
suggesting this is a fundamental characteristic of radio
galaxies and energy transport.
A small fraction of these jets are seen in the optical.
This represents a potentially profound insight into the nature
of jets since the presence of optical synchrotron radiation
demands relativistic particles of high energy and short lifetime,
unlike the radio, and in-situ acceleration processes are
probably required.

It is widely believed that a rotating black hole at the nucleus,
coupled to an inner accretion disk is somehow responsible
for the onset of a jet. The extraordinary collimation is likely to occur
at the very nucleus, though galactic influences at larger distances
may play a role in collimating the jet, \cite{junor99}.
It is also generally accepted that
both orientation (the angle of the emerging jet and of the inner accretion
disk to the line
of sight) and relativistic beaming, play roles in the appearance
of any particular source to the observer.
By such means, FR-I (or low power) radio galaxies may, when viewed from the appropriate
direction, appear as luminous, variable BL Lac objects, \cite{urry95}.

Optical emission from jets, in the beaming scenario in which the jet
moves relativistically towards the observer, is favoured
because both the typical synchrotron
spectral break is blueshifted closer to the optical window, 
and also because the intrinsic
intensity ``beam pattern'' is strongly forward projected,
with an opening angle of order $1/\gamma$ where $\gamma$
is the relativistic factor, $\gamma = \sqrt{( 1/ (1 - v^2/c^2))}$.

Other physical processes may play a role, however.
The high-pressure environment
close to a galaxy nucleus could quench the progression of
the jet and the high pressure may translate to high synchrotron
losses and a greater tendency for optical radiation. The
case of 3C~264 offers evidence of such an influence, as there the
optical jet terminates at the very edge of an inner dust lane, \cite{baum97}.
Age, or rather youth, may also play an important role. \cite{best96}
show the evolution of optical morphology of radio galaxies 
with increasing radio source size, for example,
and discuss this in the context of jet-induced star formation
and the alignment effect.

Here, we make a very simple observation that we believe strongly
supports the beaming and orientation paradigm, which is that
the majority of nearby radio galaxies showing optical jets
have compact well-ordered dust disks {\it that are almost circular}
and are hence likely to be face-on.
This implies that jets emerge close to perpendicular
to such disks and that they are pointing almost towards us.
We also note that the nuclear point source luminosities are
consistent with
this general behaviour.
In \S~2 we describe the data, in \S~3 we discuss and conclude.

\section{HST imaging observations of 3CR galaxies}

In a series of proposals, we acquired multiwavelength
images of 3CR radio galaxies and quasars using the Wide Field and
Planetary Camera 2 (WFPC2) on-board the Hubble Space Telescope (HST).
The observations utilised the efficient snapshot mode of observing, whereby
short exposures are used to fill scheduling gaps.
The data and observations are described in \cite{dekoff96}, \cite{mccarthy97},
\cite{martel98b}, \cite{lehn99}, \cite{martel99}.
An early goal of our observations  was to use the high sensitivity and
spatial resolution of HST to locate new optical synchrotron jets
in radio galaxies and hence understand the systematics
relating high energy synchrotron emission with the host properties.
That effort has been successful, and is ongoing. New optical jets were
found in 3C~78, \cite{wbs95}, 3C~15, \cite{martel98a} and elsewhere, \cite{wbs96}.
To date, there are about 12 extragalactic radio sources with known
optical jets, and a similar number of candidate optical jets.

The majority of nearby confirmed optical jets are in FR-I galaxies
and for
the purposes of this paper we define a sample
of nearby FR-I galaxies
by restricting attention to
3CR radio galaxies from \cite{benn62} whose redshift $z < 0.1$
and $|b^{II}|> 10^{\circ}$, \cite{spin85}.
Galaxies classified as either FR-I or FR-I/II
in \cite{zb95} were selected,
resulting in a sample of 25 radio sources, counting 3C~75 as two,
as listed in Table~1.
Five have known optical jets: 3C~15, 3C~66B,
3C~78, 3C~264, and 3C~274 (M87).

Nuclear point sources are common in our high resolution
images of radio galaxies, and may provide additional insight
into the beaming hypothesis.
We also therefore measured the nuclear core brightness where there
was an obvious point source using aperture photometry and the magnitude
zeropoint from the HST Data Handbook. The magnitudes are on the Vega system
and are presented in Table~1.

\section{Dust lanes in nearby 3CR FR-I radio galaxies}

\cite{dekoff00} investigated the systematics of dust lanes in
3CR galaxies from the snapshot observations and show that
in FR-I galaxies, dust lanes tend to be regular and compact
and that in such sources there is a strong tendency for the radio axis
to be perpendicular to the dust disk.
Typical scale lengths for the dust disks are hundreds of parsecs.
\cite{martel00} present an in-depth analysis of the dust-lanes
in several of these nearby radio galaxies.

\cite{baum97} presented a detailed study of 3C~264 and showed its
remarkable circular dust disk, with the bright optical jet apparently
terminating at the edge of the disk, Fig.~2.
The morphology of this dust lane was quite unique, with no other dust
lanes of the \cite{dekoff00} sample having such a circular appearance.
However, detection of round, low optical depth dust features can
be difficult since they can be significantly harder to see than
edge-on features that cut into the galaxy isophotes.

Motivated by the appearance of the 3C~264 disk, and a similar
one in 3C~66B, \cite{wbs00}, we inspected all the
remaining low$-z$ FR-I optical jet galaxies to search for the presence
of subtle dust features, and in addition those of the well-defined sample of 25 (above)
which
had not already been investigated.
This was done in case, by looking harder at the jet galaxies than the others, we were biassing
our result.
{\it However, of the 25 galaxies, 24 have clearly detectable dust.}
Even the remaining one, the Southern component of 3C~75,
shows a hint of a very faint feature in an archival HST image.
The
dust detection rate is 96\%\ and
we conclude therefore that we are not witnessing the effects of
missing subtle dust features in galaxies: our dust detection
rate is essentially complete.
Fig.~1 shows dust images of those galaxies not shown in
the references above.

Fig.~2 shows the dust in the optical jet galaxies shown in a variety of ways.
All five of the optical jet objects have multicolor HST images available.
For 3C~264 and 3C~66B no special processing is needed, and we
simply show a single direct image for each.
For the others, 
we fitted isophotes to the red
F702W galaxy images with iterative sigma
clipping to reject isophotal outliers (i.e. the jets).
The isophotal profiles were then used to make a smooth,
perfectly elliptical model of the red
light in the galaxy (where the effects of dust are minimized),
and the bluer data is divided by this red model as in \cite{wbs85}
to generate an image of the dust.
These are the $(V-R_{\hbox{model}})$ images, or ``dust images''
referred to elsewhere in the text.

The center row of Fig.~2 shows the galaxy hosting 3C~15. To the left
is a direct image, whose irregularities are attributable to dust.
In the center is a pure $(V-R)$ image, and on the right is the
$(V-R_{\hbox{model}})$ image.
The large scale circular features in the $(V-R_{\hbox{model}})$ image
are most likely artifacts from the modelling,
unlike the
more compact feature towards the center which also shows in the $(V-R)$ data image
(i.e. the ratio of the $V$ image to the $R$ image, as opposed to the ratio of the
$V$ image to a smooth, elliptical model of the $R$ image).
The dust optical depth is low
(as might be expected if the dust lane is close to face-on), but
by comparing the different images, and perhaps most easily
seen in the center one, there is a faint, almost complete, ring of absorption
and reddening, rather deeper to the West.
The jet is on a much larger scale than the dust, as it is in 3C~66B.

The lowest row of Fig.~2 shows the dust in 3C~78.
To the left is a direct $V$ image, in the center is a direct $R$ image and
on the right, the $V$ image divided by a smooth model of the
$R$ image. Again, by comparing
amongst the images, the absorption can be seen in the color image and in the direct $V$.
It is almost absent in the $R$ image. Once more, absorption can
be traced entirely around the nucleus in a compact ring.

Hence, remarkably, four of the five optical jet galaxies show circular dust
lanes.
The other one is M87  whose dust shows a more chaotic morphology, with distinct,
roughly radial filaments that coincide with line-emitting gas, \cite{wbs93}, \cite{wbs99}.
{\it None} of the other 25 galaxies show similar circular dust
lanes.

To obtain a better estimate
of the axial ratios of each of the ``round'' dust lanes,
four cuts were made at $45^{\circ}$ angles to
one another across the clearest images. The diameters were marked by eye,
and these measurements were then used to determine the parameters of the least-squares best-fit
ellipse.
The resulting measurements are given in Table~1.
Note that uncertainties tend to bias the measurements to smaller $b/a$ values
if the disk is truly circular.
The remaining dust lane axial ratios were taken either from the literature
as cited, or measured from the direct images referenced using a major and minor axis cut.

Fig.~3 shows the axial ratio distributions of the dust lanes in radio galaxies
for the 15 galaxies with reasonably well-defined dust disks,
and it is clear that the optical-jet objects have a significantly rounder distribution.
Formally, the difference is significant at the 0.1\%\ level
(i.e. a probability of chance occurrence of less than 0.001), based on the 
plotted axial ratio distributions and using the Fisher exact probability test.
The diameters of the dust disks (or rings) are approximately 1300, 350, 430 and 620~pc
for 3C~15, 66B, 78 and 264 respectively, assuming $H_0 = 75$~km/s/Mpc.

\section{Discussion}

\subsection{Dust and jet properties}

The simplest way to understand this result is that a round dust disk is
a dust disk seen close to face-on. Since we have shown in
\cite{dekoff00}, that radio jets tend to lie perpendicular to dust
disks, especially in the settled FR-I objects, the natural inference
therefore is that we are also seeing the optical jets almost pole on, that
is pointing directly towards us.
\cite{dekoff00} found that for FR-I sources
with dust disks of size $< 2.5$~kpc, the radio jet lies within less than or
about 15~degrees of
the disk (perpendicular) axis (but with a small number of exceptions such as 3C31).
The roundest disk in the sample without jets would imply an inclination $\phi$ of
order 40~degrees to the line-of-sight if it is intrinsically circular.
Similarly, the flattest disk in the jet sample would require an angle $\phi$ to
the line of sight of $\phi \approx 30$~degrees, taken at face value.
This suggests that the dispersion in angles $\phi$ is dominated by the
projection of the assembly of dust disk plus jet, rather than by
an intrinsic scatter of jet angles to dust disk angles.
It also implies that there is a critical angle, $\phi_c$, such that if
jets point within that angle we see optical emission.
From these numbers we expect $\phi_c$ to lie in the range 30---40~degrees.

There are some small asymmetries and irregularities apparent in the
circular dust lanes, and these may be due to the effects of what inclination
there is to the plane of the sky, or else to intrinsic irregularities
within the disk. Detailed analysis of the optical depth around the disk
becomes highly model dependent as there is nowhere to obtain the unobscured
light profile.

{\it Why} jets and dust lanes are close to perpendicular remains a mystery.
Since these are relaxed, orderly systems, perhaps the disk, galaxy and black hole
spin axes have had time to co-align, \cite{nat98}. Alternatively, the jet ejection process
itself may react back on
the disk and force it into a perpendicular orientation, \cite{quillen99}.
Also, in these round dust disks, the outer edge of the disk
is remarkably sharp. Perhaps this is an indication of a more settled and
evolved disk than the irregular morphologies often seen, \cite{dekoff00},
which would in turn suggest that evolutionary processes within the
jets are not responsible for the optical emission.
One would expect galaxy dynamical timescales to be much longer
than the evolutionary timescale for a radio jet (given its
high speed and short synchrotron loss lifetime).

If the sample is a set of randomly oriented linear jets, we expect the
distribution of $\phi$ to be $sin(\phi)$ for $\phi$ in the range 0 to 90~degrees.
Hence, statistically we would expect a fraction $1 - cos(\phi_c)$ to
have an angle to the line of sight $\phi$ less than a critical angle
$\phi_c$.  If $\phi_c \approx 40$~degrees, then the fraction is
23\%\ which for a sample size of 25 means six objects (three if $\phi_c = 30$~degrees).
This is obviously
consistent with the five that we find, if the optical jet galaxies
represent those whose jets point within 30---40~degrees to the line of
sight.

It is possible that more face-on disks tend to appear more irregular
and hence are excluded from the statistical comparison.  There are no
compelling reasons to think that this is actually the case, however.
One of the 10 galaxies with ``irregular'' dust has an optical jet; by
chance we would expect 1.8 galaxies to be oriented with $\phi <
35$~degrees so there is nothing inconsistent in the statistics to
suggest a large fraction of the irregular dust disks are actually
face-on.

Hence, we are led to a picture where optical jets ``appear'' when they
point to within (of order) $\phi < \phi_c \approx 35 \pm 5$~degrees to
our line-of-sight, based on the apparent axial ratios of their associated dust lanes.

This is reasonable qualitatively, given expected relativistic beaming and Doppler
boosting of the underlying synchrotron break frequency.
$\gamma$ values of order `a few' are 
typically thought to occur jets in radio galaxies, and indeed superluminal proper motion is observed
in the jet of M87 which is very direct evidence for $\gamma$ values of order $\gamma~\sim 6$,
\cite{bir99}.
If the beam angular width is $\sim 1/\gamma$, a critical angle of $\approx 40$~degrees
implies a rather lower value, $\gamma \approx 1.4$ or $v \approx 0.7 c$.
If $\phi_c = 30$~degrees, the corresponding $\gamma = 2$ and $v = 0.85 c$.
Such numbers should be taken as illustrative: clearly, in principle, beams
are observable outside the primary beam width, at much lower surface
brightness. Much deeper observations with fainter surface brightness
limits may help to disentangle the effects of multiple velocity
components and intrinsic beam pattern.

\cite{laing99} discuss the case for relativistic beaming in FR-I radio
galaxies in detail.  Based on comparison with radio data, primarily,
they develop a consistent model in which jets are intrinsically
symmetric, initially relativistic but decelerating, and faster on-axis
than at the edges.  This is quite consistent with our inference of jet
$\gamma > 1.4$ provided that either optical emission becomes
unobservable at lower values of $\gamma$ (where the beam is broad
enough for many more jets to be included within it) or else the jets
never have velocity less than $v = 0.7 c$.  \cite{laing99} finds a
best fitting velocity for single-velocity models, of $v_0= 0.72 c$, but
they also achieve good fits with lower minimum velocities when a range
of velocities is allowed.
It seems more plausible that the optical emission becomes undetectable
(with current observations) at lower velocity.

\subsection{Nuclear core sources}

Nuclear point sources are common in radio galaxies, and are often apparently
bright in galaxies with optical jets, \cite{martel98a}.
\cite{chia99} and \cite{hard00} showed that the optical core source luminosity
is very well correlated with the radio core flux, and hence is most likely
also synchrotron emission arising from the base of the jet.
In a pilot study of five objects \cite{cap99} showed that there are
indications of a correlation between gas disk inclination and core
prominence, consistent with the expectations of the proposed unification
of BL Lac and FR-I radio galaxies.
To take this a step further, we plot the nuclear magnitude where there
was obviously a point source against the axial ratio of the dust lanes,
Fig.~4. There is a trend, with the brightest cores in the most face-on disks,
and a correlation coefficient of $-0.83$ yielding a formal significance level of
better than 1\% . However, many targets do not appear on the plot. Either they
have cores but irregular dust, including two of the brightest,
or else they do not have well-defined cores in our data. We consider the
plot therefore primarily indicative and intriguing for future study.

To examine the FR-I/BL Lac unification idea, we include on
Fig.~4 the average of the BL~Lac objects from the comparison
sample of \cite{cap99}, showing the $\pm 1\sigma$ dispersion in their
magnitudes.
We assume this representative BL~Lac object is
viewed directly along the line-of-sight.
If the average spectral index is $\alpha \approx 1$ in the optical, the relativistic
boosting for continuous outflow
is $\propto (1 - \beta cos\theta)^{-2-\alpha}$, where $\beta = v/c$
and $b/a = cos\theta$.
For illustration, the best-fit
value for $\gamma \approx 5.6$ is shown, with the zeropoint of the proportionality
chosen so that the curve passes through the BL~Lac point and assuming
that all nuclei have the same intrinsic brightness.
The derived $\gamma$ is sensitive to $\alpha$, with lower $\alpha$ yielding higher $\gamma$.
While there are major uncertainties and selection effects (we only see the most luminous
cores; there is no expectation that all cores have identical intrinsic flux)
it is interesting to note that there is a good consistency
with conventional beaming models and relativistic factors between 1 and 10.
A more detailed investigation into these issues would be useful.

\subsection{Caveats}

Perhaps the most worrisome objection
to the scenario described in \S~4.1
is that in the case of 3C~264, the optical jet flares
and essentially ends {\it right at the edge of the dust disk}. This
leads one to suppose either (\romannumeral1)~a causal connection, as
discussed by \cite{baum97}, or (\romannumeral2)~a coincidence. None of
the other optical jets show any sign of interaction with the host
disks: for the most part they are on a much larger spatial scale.
In the scenario where the disk is face-on and the jet almost perpendicular to it,
the coincidence of the jet flaring out at the edge of the disk must
be accepted. In the contrary situation, where the jet ends at the edge
of the disk due to a physical interaction with the disk, the alternate coincidence that
of all galaxies, this one has the most face-on disk must be accepted.
More exotic options, such as dust spheres in the centers of optical-jet
galaxies rather than disks, can be excluded from detailed analysis
of the color distribution in the case of 3C~264, \cite{martel00}.

It may be possible to resolve this issue, either with increased sample
sizes for the optical jet cases, although that seems unlikely given our
100\%\ detection rate for nearby FR-I dust lanes, while more distant
objects are much harder to resolve, or else by demonstrating
perhaps through spectroscopic observations, that there is a
genuine interaction between the 3C~264 dust lane and the outward
propagating jet. In the latter case, we would favour a scenario in which
more than one physical process can be relevant, however it would remain very
curious as to why that one disk, of all disks, is so close to circular!

\section{Conclusions}

We have derived a simple but appealing result:
those galaxies with optical jets that are sufficiently nearby to
allow a detailed view of the nuclear environments and whose nuclei are not
so luminous as to swamp that view, show in 4 out of 5 cases evidence for a circular dust disk.
None of the other 21 galaxies in the sample show such a feature.
We interpret this as support for a two-fold model in which jets
emerge perpendicular to dust lanes and optical emission arises
when those jets are pointing towards us, specifically within a critical angle
of $\phi_c \approx 30 $---$ 40$~degrees.
Nuclear core luminosities show a trend with more luminous cores
in the face-on disks, as expected from the beaming scenario.

\acknowledgements

Support for this work was provided by NASA through proposal number
GO-5476.01-93A from the Space Telescope Science Institute, which is
operated by the Association of Universities for Research in Astronomy,
Incorporated, under NASA contract NAS5-26555.

%
%

\clearpage

\begin{table*}
\begin{center}
\begin{tabular}{lrlllll}
\tableline
3CR &  z   &    FR    &   F702W nuc mag  &             &         b/a &  Reference \\
\tableline

15  &  0.073  &  I/II  &          &      disk             &         0.92 & new \\

66B &  0.022  &  I     &   18.92  &      disk             &         0.94 & new \\
78  &  0.029  &  I     &   17.10  &      disk             &         0.88 & new \\
264 &  0.021  &  I     &   18.07  &      disk             &         0.94 & deKoff \\
274 &  0.004  &  I     &   16.10  &      irregular        &            &          \\
    &         &        &          &                       &            &           \\
29  &  0.045  &  I     &   21.0   &      irregular        &            & new \\
31  &  0.016  &  I     &   19.44  &      disk             &       0.75 & deKoff \\
40  &  0.018  &  I/II  &          &      disk             &       0.29 & Martel \\
75  &  0.024  &  I     &          &      disk in N component &    0.16 & new \\
    &         &        &          &      nothing in S     &            &        \\     
76.1&  0.032  &  I     &          &      disk             &       0.47 & new \\
83.1&  0.025  &  I     &          &      disk             &       0.09 & deKoff \\
84  &  0.018  &  I     &   15.65  &      irregular        &            & deKoff \\
88  &  0.030  &  I     &          &      irregular        &              & deKoff \\
270 &  0.006  &  I     &   19.64  &      disk             &       0.42 & deKoff \\
272.1& 0.004  &  I     &   18.30  &      irregular        &            & deKoff \\
296 &  0.024  &  I     &   21.65  &      disk             &       0.29 & deKoff \\
317 &  0.035  &  I     &   19.42  &      irregular        &            & new \\
338 &  0.030  &  I     &   20.32  &      irregular        &            & deKoff \\
353 &  0.030  &  I/II  &          &      disk             &       0.6  & deKoff \\
386 &  0.018  &  I     &   15.50  &      irregular        &            & new \\
442 &  0.027  &  I/II  &   21.12  &      irregular        &            & new \\
449 &  0.017  &  I     &   19.74  &      disk             &       0.69 & deKoff \\
452 &  0.081  &  I     &          &      disk             &       0.27 & deKoff \\
465 &  0.030  &  I     &   19.61  &      disk             &       0.73 & Martel \\
\tableline
\end{tabular}
\end{center}
\end{table*}

\clearpage

\figcaption{Dust lanes in 
galaxies not shown in previous references.
Top row 3C~29, 3C~75 (North) and 3C~76.1 from left to right respectively;
Bottom row, from left to right, 3C~317 (Abell 2052), 3C~386, and 3C~442.
Each of images on the top row is 2.3~arcsec on a side, and on the bottom row,
13.8, 6.9 and 6.9~arcsec respectively, left to right. 3C~75 and 3C~76.1
are simple, direct images, while the remainder are $(V-R_{\hbox{model}})$ images. 3C~29 shows a remarkable X-shape; 3C~317 a chaotic extensive dust distribution
(the giant elliptical galaxy would be centered); 3C~386 (which has a very bright point source) faint patches to the West
and North-East, and 3C~442 an irregular fan to the North West.}

\figcaption{Montage of dust disks in nearby radio galaxies with optical jets.
Top row, from left to right, 3C~264 (3.3~arcsec on a side) F547M direct image; 3C~66B
(3.4~arcsec) FOC F430W direct image, and 3C~274 (11.9~arcsec) from \cite{wbs99}.
Center row shows 3C~15, 2.9~arcsec on a side, from left to right direct $V$,
$(V-R)$ and $(V-R_{\hbox{model}})$ respectively.
Bottow row shows 3C~78, 2.3~arcsec on a side, from left to right direct $V$,
direct $R$ and $(V-R_{\hbox{model}})$ respectively.
\label{fig2}}

\figcaption{Dust axial ratio distributions for dust lanes with disk-like morphology, data from Table~1.
The shaded region indicates those sources with optical jets.\label{fig3}}

\figcaption{Core source luminosity versus dust axial ratio. The galaxies with
optical jets are shown as `$+$' signs, the other
radio galaxies as open symbols, and for comparison, we include the average point
for the BL~Lac comparison sample of \cite{cap99} (the cross).
The curve is a beaming model assuming fixed total core luminosity and $\gamma = 5.6$.
\label{fig4}}

\clearpage

\plotone{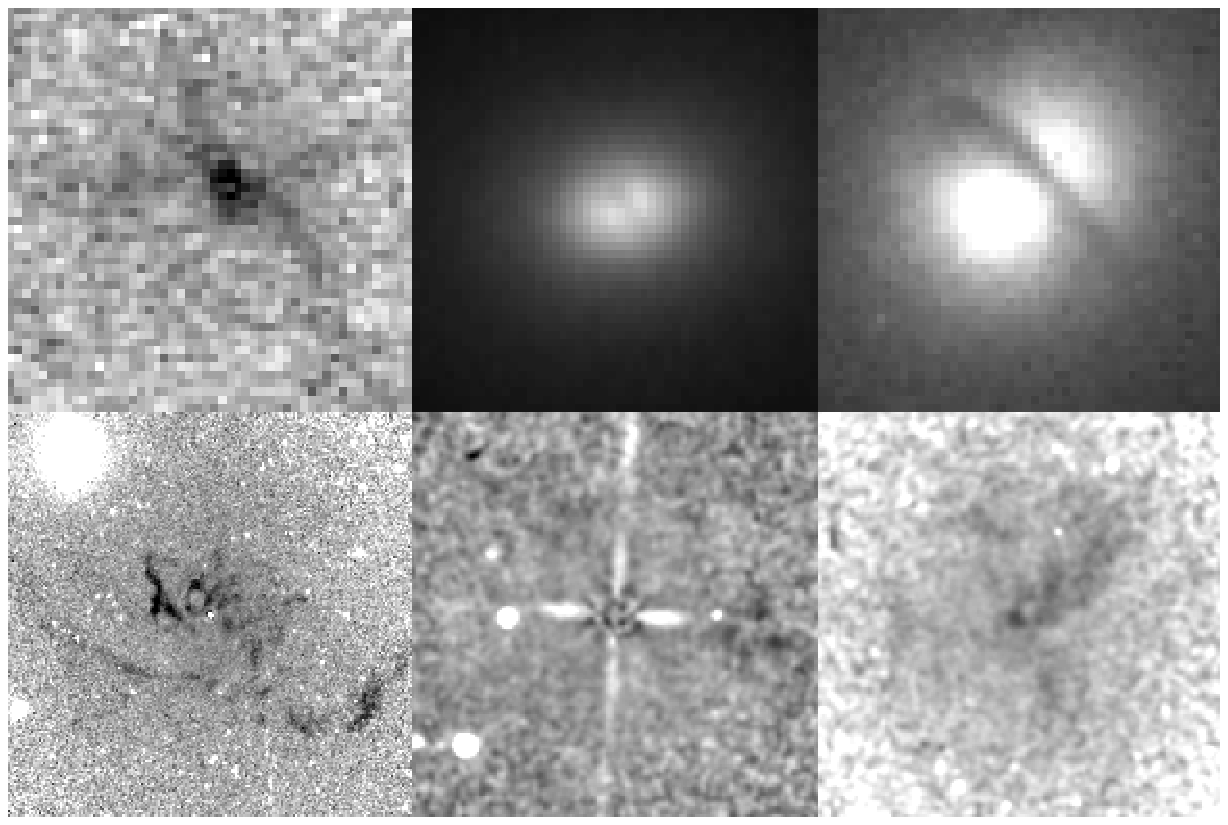}

\clearpage

\plotone{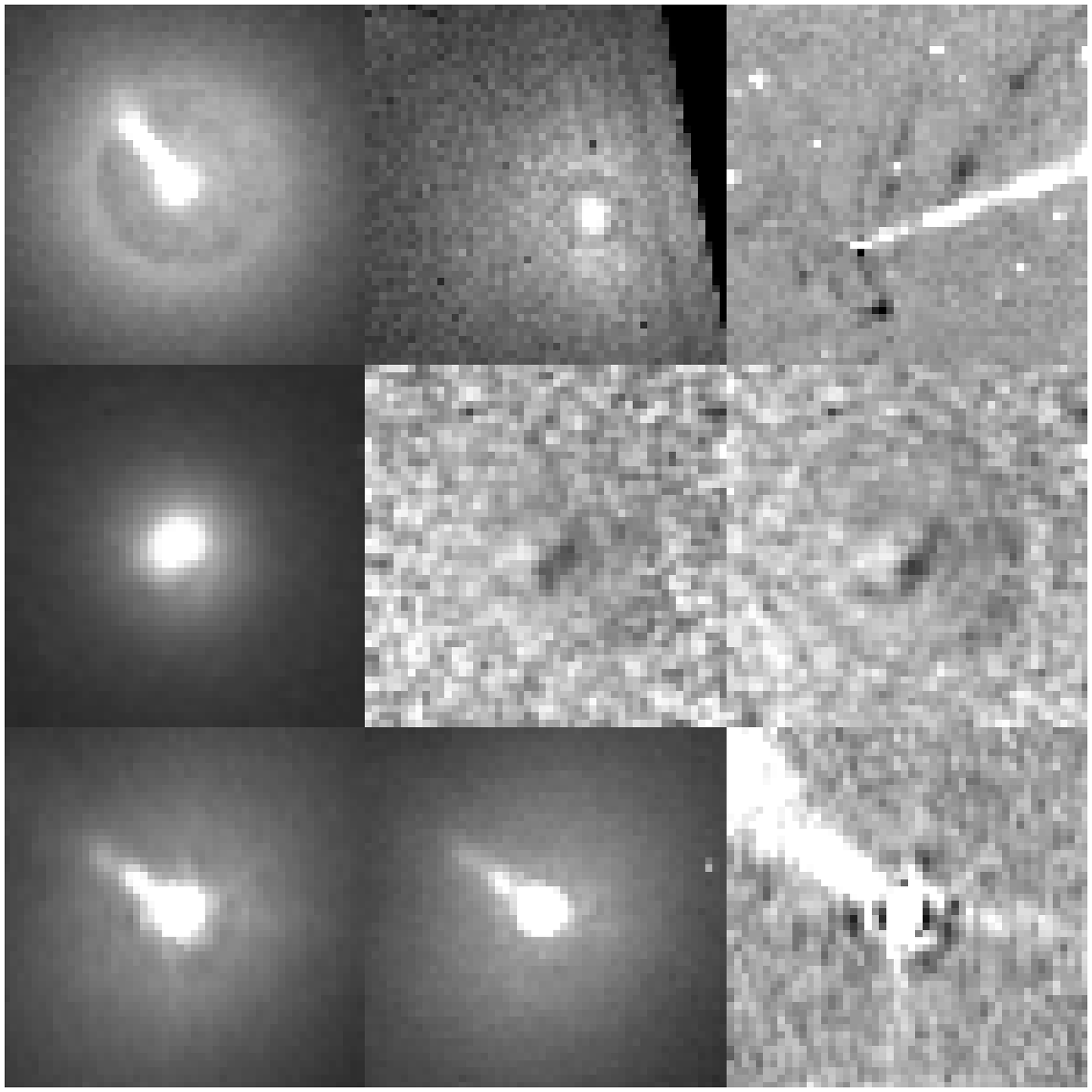}

\clearpage

\plotone{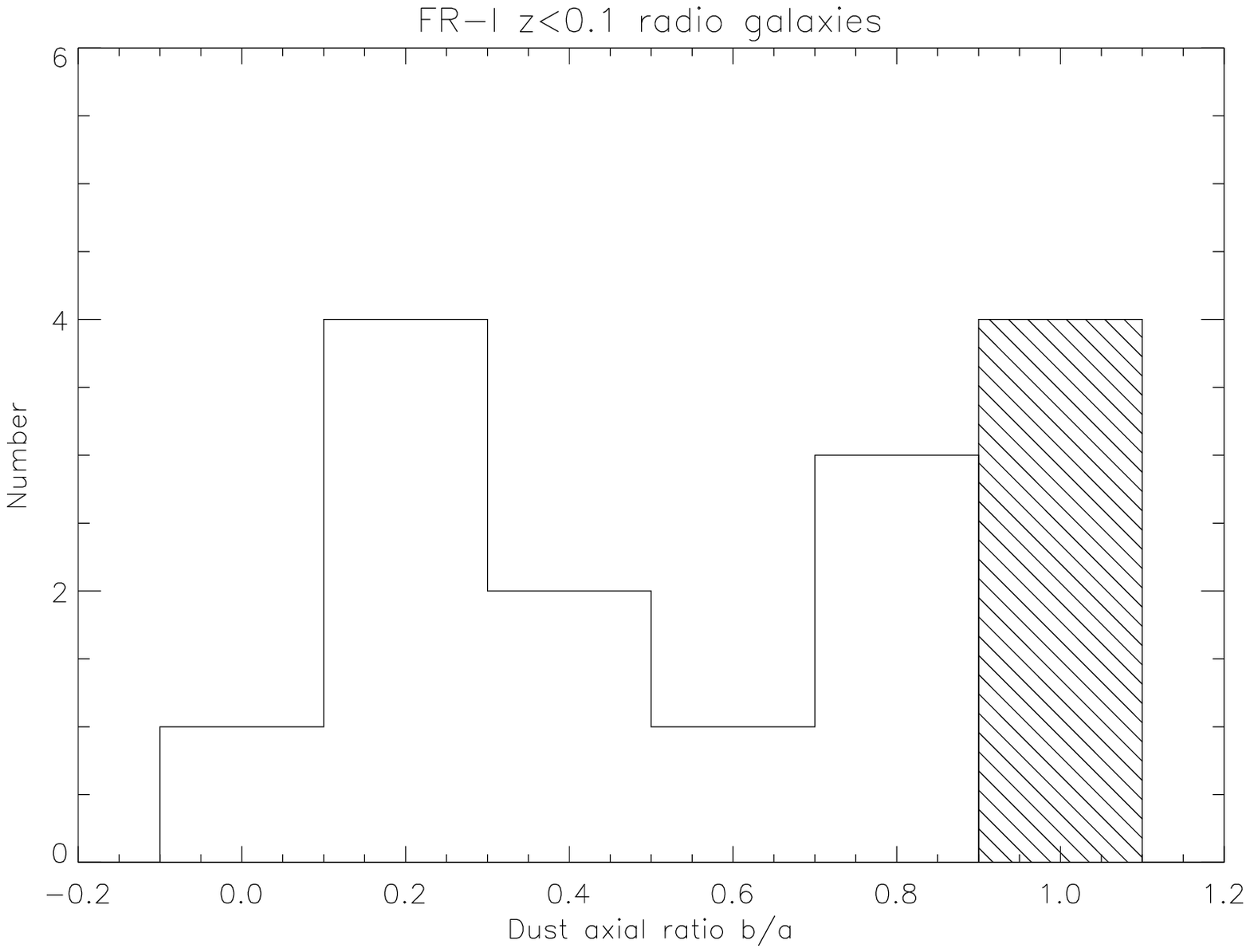}

\clearpage

\plotone{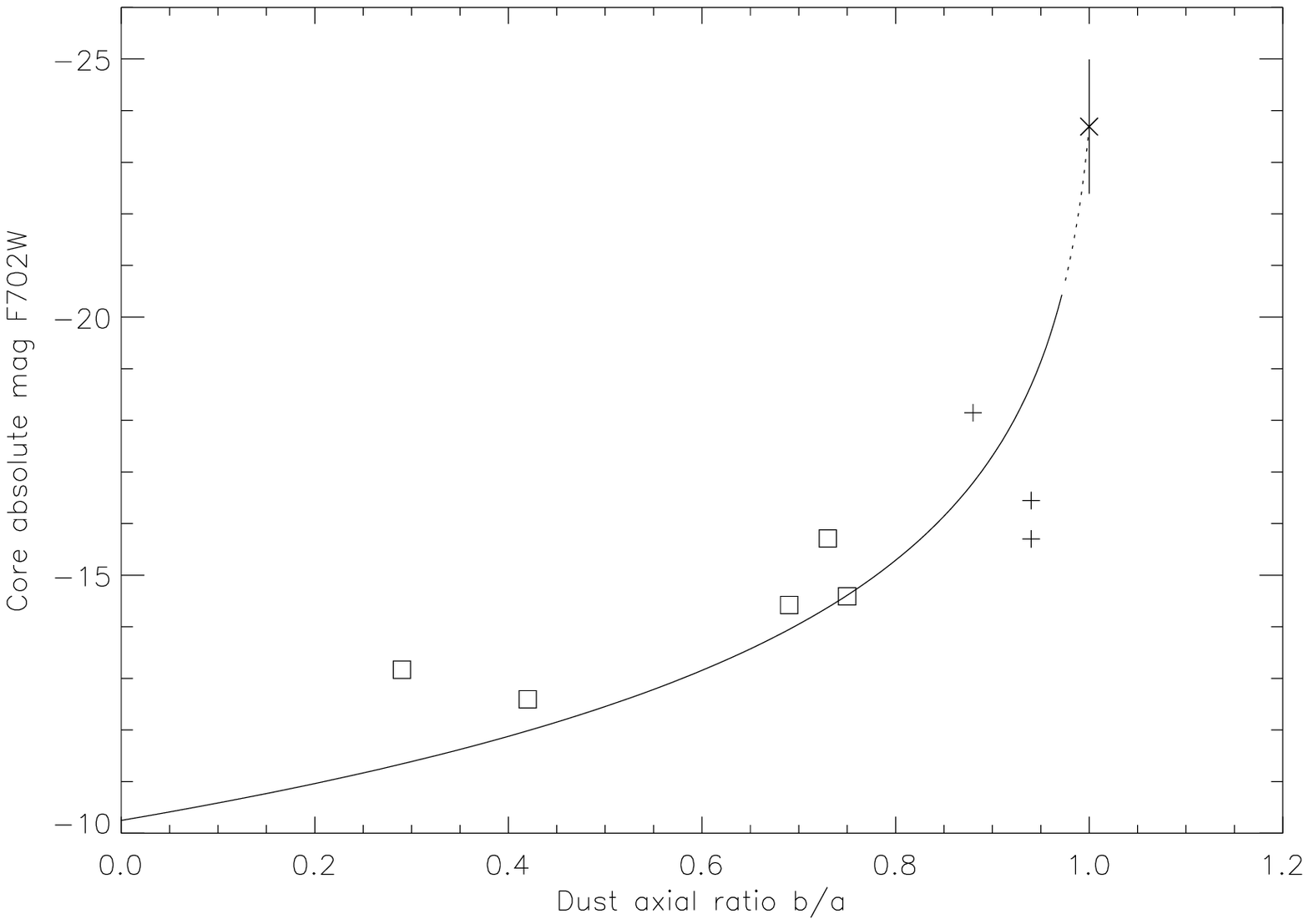}

\end{document}